\documentclass[traditabstract]{aa} 
\usepackage{graphicx}
\usepackage{txfonts}
\usepackage{color}
\newcommand{\kms}{km s$^{-1}$}
\newcommand{\teff}{$T_{\rm eff}$}
\newcommand{\logg}{$\log g$}
\newcommand{\turb}{$\xi$}
\newcommand{\vsini}{$v \sin i$}

\newcommand{\msol}{M$_{\odot}$}
\newcommand{\rsol}{R$_{\odot}$}
\newcommand{\lsol}{L$_{\odot}$}
\begin{document}
   \title{HD~172189: another step in furnishing one of the best 
laboratories known for asteroseismic studies}


   \author{O.~L.~Creevey
          \inst{1,2}$^{\star}$,
          K.~Uytterhoeven\inst{1,3},
          S.~Mart\'in-Ruiz\inst{4},
          P.~J.~Amado\inst{4},
          E.~Niemczura\inst{5},\\
          H.~Van~Winckel\inst{6}
          J.~C.~Su\'arez\inst{4},
          A.~Rolland\inst{4},
          F.~Rodler\inst{1,2},
          C.~Rodr\'iguez-L\'opez\inst{4,7,8},
          E.~Rodr\'iguez\inst{4},
          G.~Raskin\inst{6,9},
          M.~Rainer\inst{10},
          E.~Poretti\inst{10},
          P.~Pall\'e\inst{1,2},
          R.~Molina\inst{11}, 
          A.~Moya\inst{4},
          P.~Mathias\inst{12},
          L.~Le Guillou\inst{6,9,13},
          P.~Hadrava\inst{14},
          D.~Fabbian\inst{1,2},
          R.~Garrido\inst{4},
          L.~Decin\inst{6},
          G.~Cutispoto\inst{15},
          V.~Casanova\inst{4},
          E.~Broeders\inst{6},
          A.~Arellano Ferro\inst{11},
          \and
          F.~Aceituno\inst{4}
          }

\institute{
  Instituto de Astrof\'isica de Canarias, 
  E-38200 La Laguna, Tenerife, Spain.
  $^{\star}$\email{orlagh@iac.es}
  \and
  Departamento de Astrof\'{i}sica, Universidad de La Laguna, 
  E-38205 La Laguna, Tenerife, Spain.
  \and
  Laboratoire AIM, CEA/DSM-CNRS-Universit\'e Paris, Diderot;
  CEA,IRFU, SAp, centre de Saclay, F-91191, 
  Gif-sur-Yvette, France.  
  \and
  Instituto de Astrof\'isica de Andaluc\'ia (CSIC), 
  Camino bajo de Hu\'etor 50,
  18080 Granada, Spain.  
  \and
  Astronomical Institute, Wroc{\l}aw University, Kopernika 11, 52-622
  Wroc{\l}aw, Poland.
  \and
  Instituut voor Sterrenkunde, Celestijnenlaan, 200D, 3001 Leuven, Belgium.
  \and
  Laboratoire d'Astrophysique de Toulouse-Tarbes, Universit\'e de Toulouse, CNRS 
  31400 - Toulouse, France.
  \and
  Universidad de Vigo, Departamento de F\'isica Aplicada, 
  Campus Lagoas-Marcosende, 36310-Vigo, Spain.
  \and
  Mercator Telescope, Observatorio del Roque de los Muchachos, 
  Apartado de Correos
  474, 38700 Santa Cruz de La Palma, Spain.
  \and
  INAF-OABrera, Osservatorio Astronomico di Brera, Via E. Bianchi 46, 23807 Merate, Italy.
  \and
  Instituto de Astronom\'ia, Universidad Nacional Aut\'onoma de Mexico, Apdo. Postal 70-264, 04510 Mexico D.F., Mexico.
  \and
  UNS, CNRS, OCA, Campus Valrose, UMR 6525 H. Fizeau, F-06108 Nice Cedex 2, France.
  \and
  UPMC Univ. Paris 06, UMR 7585, Laboratoire de Physique Nucl\'eaire et
  des Hautes Energies (LPNHE), F-75005 Paris, France.
  \and
  Astronomical~Institute, Academy~of~Sciences,
  Bo\v{c}n\'{\i}~II~1401, CZ~-~141~31~Praha~4,
  Czech~Republic.
  \and
  INAF Catania Astrophysical Observatory,
  via S. Sofia, 78-95123, Catania, Italy. }

   \date{Received September 15, 1996; accepted March 16, 1997}

  \abstract
     {HD~172189 is a spectroscopic eclipsing binary system with a 
        rapidly-rotating pulsating 
        $\delta$ Scuti component.  It is also a member of the open cluster 
        IC~4756.  These combined characteristics make it an excellent
        laboratory for asteroseismic studies.
      To date, HD~172189 has been analysed in detail photometrically but not 
        spectroscopically. For this reason
        we have compiled a set of spectroscopic data 
        to determine the absolute 
        and atmospheric parameters 
        of the components. 
We determined the radial velocities (RV) of both components using 
        four different techniques.
     We disentangled the binary spectra using KOREL, and performed the
        first abundance analysis on both disentangled spectra.
        By combining the spectroscopic results and the photometric data,
        we obtained the component masses,
        1.8 and 1.7 \msol, and radii, 4.0 and 2.4 \rsol, for 
        inclination $i = 73.2^{\circ}$, 
        eccentricity $e = 0.28$, and orbital period $\Pi = 5.70198$ days.
        Effective temperatures of 7600 K and 
        8100 K were also determined.
        The measured 
       $v \sin i$ are 78  and 74 \kms, respectively,
        giving rotational periods of 2.50 and 1.55 days for the components.
        The        
        abundance analysis shows 
        [Fe/H]~=~--0.28 for the primary (pulsating) star, consistent
        with observations of IC~4756.
        We also present an assessment of the different analysis techniques 
        used to obtain the RVs and the global parameters.}


\authorrunning{Creevey et al.}
\titlerunning{HD~172189: global parameters of a $\delta$ Scuti star}
\maketitle

%
%
%

\section{Introduction}

    HD~172189 (=BD +5 3864, $V = 8.85$~mag, 
    $\alpha = 18^{\rm h}$38$^{\rm m}$37.6$^{\rm s}$, 
    $\delta = +05^{\rm d}27^{\rm m}55.3^{\rm s}$, J2000) 
    has the combined characteristics of being an
    {\it eclipsing} and {\it spectroscopic} binary, {\it pulsating star}, and 
    {\it member of a cluster}
    (Martin 2003, Mart\'in-Ruiz et al. 2005 --- MR05 hereafter, 
    Costa et al. 2007, Ibano\v{g}lu et al. 2009 --- I09 hereafter).
    Each of these provide unique constraints that allow
    us to test stellar evolution theories in an independent form:
    a) an eclipsing spectroscopic binary system is fundamental for 
    determining the 
    absolute global parameters of both stars and the system with precision; 
    b) a pulsating star allows us to use the oscillation frequencies to probe
    the interior of the star, thereby also determining the 
    evolutionary state;
    c) cluster membership has the distinct advantage that the properties such
    as age, metallicity, and distance can be well-determined.
    Given the constraints imposed by the cluster membership and the 
    binary system on the mass, age, and metallicity of the pulsating star,
    the observed seismic frequencies  can be used 
    to test and improve the current asteroseismic models.
    For example, as both components are rapidly rotating 
    with periods of 2.50 and 1.55 days, (see Sect.\ref{sec:photo}),
    we can 
    investigate the effects of rotation, such as the mixing of elements 
    and transport of angular momentum.
    Several theories exist regarding rapid rotation, but as yet, 
    observations have not been able to confirm any of these hypotheses.
    Some examples of these unproved theories include understanding
    the interplay between rapid rotation and convective cores, enabling
    the transport of angular momentum both poleward and in the radial
    direction (Featherstone et al. 2007), 
    or the existence 
    of an overshoot boundary layer between the convective core and the 
    radiative region, where mixing of nuclear elements can influence
    main sequence lifetimes (Brun et al. 2004).
    Such theories can be confirmed from detailed seismic modeling,
    once the global parameters of the pulsating star have been determined.

    Since HD~172189 was discovered to be a binary system
    by Martin (2003), 
    several groups have shown a keen interest in this object.
    Dedicated photometric campaigns (Amado et al. 2006, MR05,
    Costa et al. 2007, I09)
    have begun to reveal the true nature of this system, 
    and several 
    oscillation
    frequencies have been documented from the time series of the 
    $\delta$ Scuti star.
    This system 
    is moreover a selected target 
    of the asteroseismic core program of the CoRoT satellite 
    mission (Baglin et al. 2006a,b, Michel et al. 2008), 
    and has been continuously observed from space 
    in white light for about 150 days from April to September 2008, 
    with the aim of interpreting the pulsations.
 
    With the prospects of using the observed oscillation frequencies to 
    study the internal structure of the star, we have compiled 
    spectroscopic data taken in 2005 and 2007 from various sources with the 
    aim of determining some spectroscopic properties of the system, to 
    facilitate the future analysis of this star.
    We determine the radial velocities (RVs) of the individual components using
    various techniques (Sect.~\ref{sec:rvs})
    to subsequently 
    solve for  the orbital parameters  of the sytem, while also
    providing an assessment of the methods employed (Sect.~\ref{sec:orbital}).  
    We combine the RV data with photometric data and 
    present
    a  full  orbital  and
    component solution for this object (Sect.~\ref{sec:photo}).
    We subsequently disentangle the spectra (Sect.~\ref{sec:korel}), 
    estimate the effective temperatures \teff\ using the 
    disentangled spectra, and using
    synthetic spectra (Sect.~\ref{sec:teff})
    and perform the first abundance analysis of this object 
    (Sect.~\ref{sec:abund}).
    Discussion and conclusions then follow.
    Before beginning our analysis, we briefly review the literature of both
    HD~172189 and IC~4756.


    One of the first references to HD~172189 and IC~4756 
    can be found in Graff (1923), where HD~172189 is named star 83 
    and has $V=8.69$. 
    Later, Kopff (1943) published an analysis 
    (star 93,  
    $V = 8.86$ mag), using a referenced tentative 
    spectral typing of A6 from Wachmann (1939).
    Photoelectric observations of IC 4756 were then carried out in 
    1964 in Lowell
    Observatory (Alcaino 1965), with the 
    purpose of 
    determining the distance and the 
    absorption of the cluster 
    as well as to establish a criterion for
    membership.
    This author determined from the measured $V$ = 8.73 mag 
    and the colour-colour
    diagram, that HD~172189 (Alcaino star 10) most likely was {\it not}
    a member of the cluster, however, stated that 
    proper motions would be needed to 
    confirm this.  They determined a distance modulus of 8.2 mag 
    corresponding to 437 pc, and an age of 820 Myr for IC 4756.
    Herzog et al. (1975) subsequently determined the proper motions
    of 464 stars in the field of this cluster 
    and estimated a probability of 89\% of HD~172189 (Herzog 205)
    being a member of the cluster, while
    Missana \& Missana (1995) obtained a 91\% probability based on
    proper motions and the position of the star.

    Schmidt and Forbes (1984) measured a \vsini~of 69 kms$^{-1}$,
    where $i$ is the inclination of the rotation axis (assumed
      to be the same for both stars and equal to the inclination of the orbital
      plane). 
    The spectral type has been somewhat discordant in the literature, 
    probably due to its binary nature later discovered by 
    Martin (2003).
    Adding the fact that spectral typing of A stars can be difficult and 
    that both components of this (at least) double system  
    (see Sect.~\ref{sec:discussion}) 
    are rapidly rotating,
    it is not a surprise that different authors
    have arrived at several inconsistent results: 
    A6 V (Wachmann 1939, Herzog et al. 1975), 
    A7 III (Schmidt \& Forbes 1984), 
    A6 III or A4 III (Dz{\v e}rvitis 1987),
    A2 V (Costa et al. 2007), and 
    A6 (I09).
    Schmidt (1978) measured Str\"omgren photometry of the system:
    $\beta = 2.820$, $(b-y)$ = 0.258 mag, $m1 = 0.123$ mag, 
    $c1 = 1.055$ mag, and estimated
    $E(b-y)$ = 0.15.  
    Using the de-reddenned quantities and the tables from Cox (2000), HD~172189
    appears to be of spectral type late B or early A. 
    The range of these spectral types clearly imposes little constraint on
    $T_{\rm eff}$, luminosity $L_{\star}$ and  gravity $\log g$.
    
    With the available photometric data of the eclipsing binary system
    (MR05, Amado et al. 2006, Costa et al. 2007), extra 
    constraints can be imposed
    on some of the fundamental parameters. 
    Very recently, I09 published combined 
    photometric and spectroscopic data with estimates of global parameters.
    They suggested that the system has component masses 
    of 2.06$\pm$0.15 and 1.87$\pm$0.14 \msol. 
    The cluster has an age of roughly 1 Gyr
    (Alcaino 1965, Mermilliod \& Mayor 1990).
    The two components are rapidly rotating in a non-synchronous fashion, 
    and the orbit is quite eccentric, indicating that the stars 
    are not interacting
    and are most likely detached MS stars.  Furthermore, 
    the suggested MS turn-off mass for IC~4756 is 1.8--1.9 \msol\
    (Mermillod \& Mayor 1990).
    
%
%
%


    \begin{table}
      \begin{center}
        \caption{ Summary of spectroscopic observations
          \label{table:obslog}}
        \begin{tabular}{lllccc}
          \\
          \hline\hline
          Observatory&Instrument&Dates&\#Spectra (Used)\\
          \hline
          La Silla &2.2m+FEROS &June-July, 2005 & 17\\
          OHP&1.52m+Aurelie & June, 2005 & 14 (12) \\ 
          Catania&0.91m+FRESCO & May-Aug, 2005 & 21 (19)\\
          Calar Alto & 2.2m+FOCES & May-June, 2007 & 11\\
          La Palma & NOT+FIES & July, 2007 & 14 (11)\\
          \hline
          \hline
        \end{tabular}
      \end{center}
    \end{table}
    


   \section{Spectroscopic Observations\label{sec:observations}}

    The spectroscopic observations used in this work are summarised in 
    Table \ref{table:obslog}, and  
    the following subsections describe the observation, 
    reduction, and calibration of the 
    spectra taken at the various observatories.
    All of the spectra were subsequently barycentric corrected.

    \subsection{Aurelie data}
    A total of 14 spectra of HD~172189 were obtained in a time span of 10 
    nights from 15 to 25 June 2005 with the Aur\'elie spectrograph, 
    mounted on the 1.52m telescope, at Observatoire de Haute Provence (OHP), 
    France. 
    The instrument has a grating of  1800 lines/mm, providing a spectral 
    resolution $R \simeq$ 25,000. 
    The spectral range used was 4528--4675\AA/4468--4542\AA, 
    with  exposure times of 1200/1500, or 3600 seconds. 
    As there is no pipe-line reduction available, we used 
    standard IRAF  (Tody 1986) reduction procedures. 
    The continuum normalisation was performed manually by fitting a 
    cubic spline. 
    Typical signal-to-noise ratio (SNR) values of the spectra are 55--63.
    
    \subsection{FIES data}
    The FIES data were obtained during an observation run of a separate object 
    in July 2007 using the 2.5m NOT telescope at the Observatorio 
    del Roque de los Muchachos.
    The FIbre-fed \'{E}chelle Spectrograph (FIES) is 
    a cross-dispersed high-resolution \'{e}chelle spectrograph.  
    We used the medium-resolution setup with $R \simeq 45,000$.
    The spectral range is 3640--7455 \AA, with a maximum 
    efficiency of 9\% at 6000 \AA.
    Separate wavelength calibrator exposures of 
    Thorium-Argon (ThAr) were obtained. 
    During each of the 5 nights of observations, 
    either 2 or 3 spectra of HD~172189 were taken with exposure times 
    of 600 seconds. 
    We used the FIEStool 
    reduction software (Stempels, 2004\footnote{Available at {\tt 
        http://www.not.iac.es/instruments/\\fies/fiestool/FIEStool.html}})
    to calibrate the wavelength
    and reduce the spectra.  
    This tool is optimised for FIES data, although 
    the reduction is standard and calls IRAF to perform some of the 
    tasks.
    SNR values around 5720 \AA\ are $\sim$70.
    
    \subsection{FRESCO data}
    From May -- August 2005 a total of 21 spectra of HD~172189 were 
    observed with the FRESCO \'echelle spectrograph, attached to the 91cm 
    telescope of the  M.~G.\, Fracastoro Mountain Station at
    Catania Astrophysical Observatory (CAO), Sicily, Italy. 
    The FRESCO spectra, with a resolution $R \simeq 21,000$, span the spectral 
    range from 4320 to 6800 \AA\, recorded on 19 orders.  
    Standard IRAF reduction procedures were used
    and the  continuum normalisation was
    performed manually by fitting a cubic spline.
    Typical exposure times were 3600 seconds and 
    SNR values near 5720 \AA\ are $\sim$45.

    \subsection{FEROS data}
    The Fiber-fed, Extended Range, \'Echelle Spectrograph (FEROS), mounted
    at the 2.2m ESO/MPI telescope at La Silla (ESO), Chile,  
    has a resolution of 
    $R\simeq 48,000$ and covers almost the complete range between
    3500 and 9200 \AA\
    on 39 \'echelle orders. A total of 
    17 spectra were taken in June-July 2005 with exposure times 
    of between 750 and 900 seconds. We reduced the spectra using an improved 
    version of the standard FEROS pipeline (see Rainer 2003). 
    We used an automated
    continuum normalisation procedure developed by M.~Bossi (INAF
    OAB-Merate) to normalise the spectra. 
    Typical SNR values in the region of  5720 \AA\ are $\sim$130.

    \subsection{FOCES data}
    Observations  at the  observatory  of Calar  Alto  in Almeria,  Spain,
    during  10 nights in May-June  2007 were  carried out  both in
    visitor and  in service mode.  
    The full optical range with a resolution of $\sim$35,000 was recorderd
    with the \'echelle spectrograph FOCES on the 2.2m telescope.
    In total,
    11 exposures of HD\,172189 were collected each with exposure
    time of 1200 seconds. 
    The \'echelle data were reduced using the standard IRAF procedures. 
    Typical SNR values 
    in the region of 5720 \AA\ are $\sim$57.



    \section{Determination of radial velocities\label{sec:rvs}}

    As both components of the  system are rapidly rotating, the spectral
    lines  are  broadened,  making  it difficult  to  identify  individual
    lines Doppler-shifted due to the orbital motion.  In
    fact,  most profiles  are merged 
    and line profile variations due to pulsations are clearly present
    making complicated and delicate the determination of the RVs.
    For these reasons, 
    we used the following independent methods to determine the RVs,
    described in the next subsections:
    \begin{itemize}
    \item LSD+GAU: Fitting a  double-Gaussian function to 
      the Least-squares deconvolution (LSD) 
      profiles (Donati et al. 1997, 1999).
      The central positions of the Gaussian functions are the RVs of
      each component.
    \item LSD+MM: Calculating the first moments (Aerts et al. 1992) 
      from the LSD profiles.
    \item IRAF: 
      Using the IRAF task FXCOR with a synthetic non-broadened
      template and then using the DEBLEND function to compute the
      RVs.
    \item KOREL: Disentangling the spectra using the programme 
      KOREL (Hadrava 1995), 
      and determining the RVs by
      fitting the observed spectra with the superposition 
      of the Doppler-shifted disentangled spectra.
      \end{itemize}


    \begin{figure}
      \includegraphics[width=0.5\textwidth]{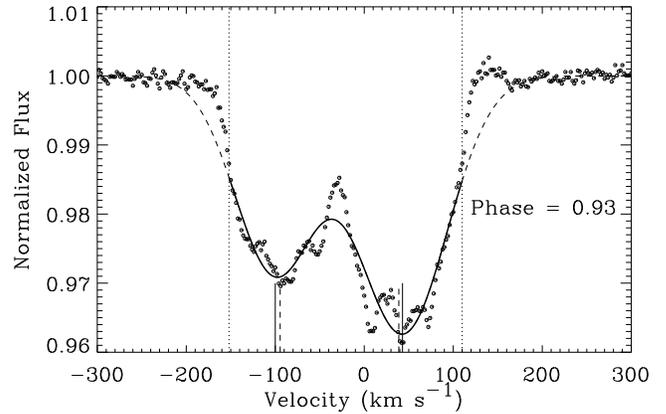}
      \caption{ Least-squares deconvolution (LSD) profile (open circles) 
        calculated 
        at phase 0.93 (based on T$_{\rm min}$ = 2452914.644 HJD from MR05) 
        by summing and weighting all
        of the lines in the spectrum.
        The solid curve is the double-Gaussian function fitted to the LSD 
        profile using just the data points within the region delimited by
        the dotted vertical lines.
        We have highlighted the extracted RVs from the 
        LSD+GAU and LSD+MM 
        methods
        with vertical continuous and dashed lines, respectively.
        \label{fig:lsdprofile}}
    \end{figure}


    \subsection{Least-squares deconvolution\label{subsec:lsd}}
    
    To make optimised use of the common information available in all
    spectral lines, we deconvolved several hundreds of individual lines by
    comparison with synthetic masks, into a single 'correlation' profile
    with a significantly increased SNR, using the Least-squares
    deconvolution (LSD) method (Donati et al. 1997, 1999) 
    (cf. Fig.~\ref{fig:lsdprofile}).  
    The synthetic line
    masks contain lines from the VALD database 
    (Piskunov et al. 1995, Ryabchikova et al. 1999, Kupka et al. 1999).
    We used the values of 
    $T_{\rm eff}=8250$ K and $\log g = 4.0$ dex (see Sect.~\ref{sec:teff}) 
    for the spectral template.
    Varying the effective temperature with 
    $\Delta T_{\rm eff}=250$ or $500$ K and the gravity 
    with $\Delta \log g = 0.05$ or 0.1 dex
    did not affect the central positions of the 
    LSD profiles significantly. 
    The fitted RV values obtained from using different templates was 
    used to estimate the error on the RVs.
    The LSD profiles were calculated by
    taking into account all elements, apart from He and H, in the regions
    4380--4814 \AA\ and 4960--5550 \AA.
    This involved deconvolving about 3000,
    2600, 2200, 1525, or 250 lines for the FEROS, FIES, FRESCO, 
    FOCES, and Aurelie spectra, respectively, and resulted in
    profiles with SNR of 
    1100--3300, 500--1700, 400--2000, 400--1200, and 200--700.
    The LSD line profiles did not all have a continuum level at 1.0, so 
    they were
    subsequently normalised, followed by a rescaling to homogenise the
    profiles obtained with different instruments (see description in
    Uytterhoeven et al. 2008).  The velocity steps of the LSD profiles
    were 2 km\,s$^{-1}$ (FIES, FEROS and FOCES), 2.3 km\,s$^{-1}$
    (Aurelie), and 7.8 km\,s$^{-1}$ (FRESCO). We note that systematic
    instrumental RV offsets of the order of $1.5$\,km\,s$^{-1}$ are most
    likely present (see Uytterhoeven et al. 2008). However,
    due to the small amount of datapoints per dataset, 
    we currently are not
    able to quantify and subsequently correct for these 
    instrumental differences.

    \subsection{Double-Gaussian fit to the LSD profile\label{sec:doubleg}}

    To determine the RVs of each component we fit a 
      double-Gaussian function
    to each LSD profile.
    We initially fixed the widths of the Gaussians while fitting the 
    central positions (the RVs) and the amplitudes, and subsequently allowed 
    all of the Gaussian parameters to be fit.
    Cut-off values in the velocity axis were used for each fit
    because the broad wings
    of the Gaussian profiles do not accurately match the LSD profiles.
    Fig.~\ref{fig:lsdprofile} shows an example of an LSD profile when 
    both components are near maximum separation at orbital 
    phase 0.93 (based on reference eclipse minimum of HJD
    2452914.644 days, and orbital period $\Pi$ = 5.70198 days from MR05).
    The deformations in the profiles are most likely due to pulsations,
    and this inhibits the accuracy of the RV.
    The dotted vertical lines
    delimit the region that was fitted.
    The solid curve shows the Gaussian fit, and the dashed curve shows the 
    rest of this function for further velocity values.   
    The model is fit several times using various cut-off values, and 
    using several LSD profiles which are calculated from different 
    spectral templates. 
    The RVs are defined as the mean values of these fits, with the standard 
    deviations defining the errors.
    The RVs corresponding to this spectrum are denoted by the 
    vertical continuous lines.
    This method worked well when the components in the LSD 
    profiles were sufficiently
    separated, hence RVs at conjunction are not available using this method.

    \subsection{First normalised moments} 

    Another tool to obtain RVs for the primary and secondary 
    components from the LSD profiles is to calculate the first 
    normalised moments ($\langle v\rangle$,  e.g. Aerts et al. 1992). 
    At quadrature the velocity profiles of 
    the primary and secondary stars are well separated. 
    In these orbital phases we determined the integration boundaries, 
    within which to
    calculate $\langle v\rangle$, from the individual Gaussian 
    profiles for the primary and secondary components 
    described in Subsect.~\ref{sec:doubleg}.
    Close to conjunction, the velocity contributions 
    of the primary and secondary stars are blended, 
    which complicates the definition of the profile boundaries. 
    Therefore, we assumed a fixed width of the component profiles, 
    derived from the spectra in elongation phase.  
    The moments were calculated by determining 
    one of the integration borders, and calculating 
      the other border assuming a fixed profile width.
    This method is sensitive to the profiles, 
    including any deformations due to 
    pulsations.  In Fig.~\ref{fig:lsdprofile} the RVs 
    corresponding to this
    spectrum are denoted by the vertical dashed lines 
    showing an offset of a few \kms\ from the values of LSD+GAU.

    \subsection{IRAF FXCOR}

    FXCOR cross-correlates the observed spectra with a template
    spectrum in the Fourier domain. In our case, the merged (one-dimensional) 
    spectra of HD\,172189 were cross-correlated with a
    Kurucz synthetic spectrum computed with the approximate physical
    parameters of the components of the binary, i.e., 
    $T_{\rm eff}=8000$~K and 7500~K, $\log{g}=4.00$ and solar abundance.
    
\begin{figure}
      \includegraphics[width=0.5\textwidth]{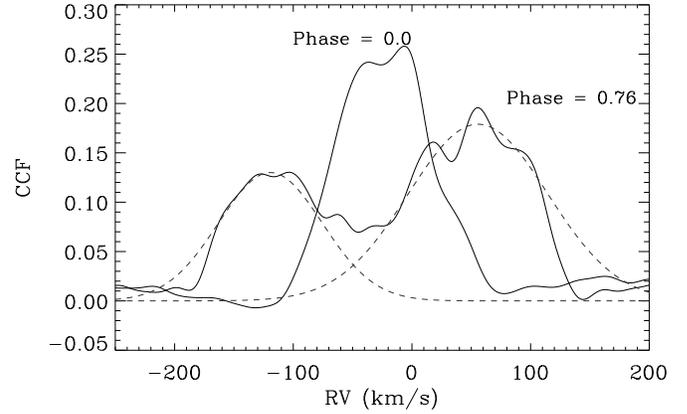}
      \caption{ Cross-correlation profiles (CCF) of the observed spectra with 
        a template spectrum at two different orbital phases (solid curves).
        The dashed curve is a double-Gaussian fit 
        at phase 0.76, obtained by fitting
        the upper 50\% of the profiles.
        \label{fig:ccfpedro}}
    \end{figure}

    The  range of  the spectrum  used for  the cross-correlation  was 
    between 5000 and 6500 \AA,  avoiding the two regions most affected
    by telluric absorption lines. The object spectra were filtered in the
    Fourier domain with a bandpass filter according to the
    resolution of the data. Most information in the Fourier spectrum is
    above a certain wavenumber which was computed to take into account the
    resolution of our data. The Fourier transform of the spectrum was then
    multiplied by a ramp function which starts rising at wavenumber 10,
    reaches 1 at 20, starts falling at 2500 and reaches 0 again at 4500. 
    The cross-correlation functions (CCF) were then calculated.
    Two examples of the CCFs at different orbital phases
    are shown in Fig.~\ref{fig:ccfpedro}.
    Gaussian functions were subsequently fitted to the CCFs,  
    using the full CCF profile (later referred to as IRAF$_{\rm FULL}$), 
    and next using only the upper 50\% of the CCF (later referred to as IRAF),
    resulting in much better central Gaussian fits.


    \begin{figure}
      \includegraphics[width = 0.5\textwidth]{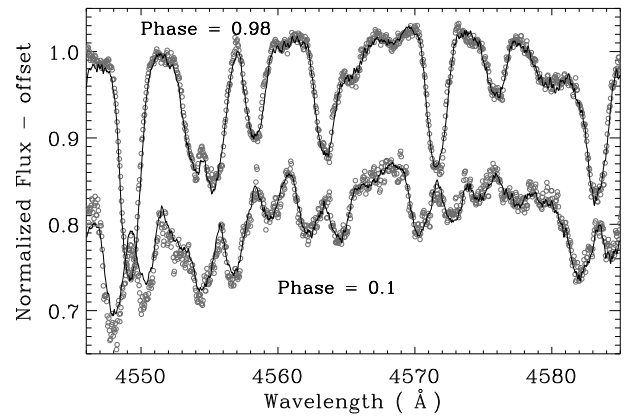}
      \caption{ Fit of the observed spectra by KOREL disentangling.
        The circles are the observations while the solid curves 
        are the superimposed RV Doppler-shifted disentangled
        spectra (y-shifted for clarity). \label{somedisp}}
    \end{figure}


    \subsection{Disentangling of spectra}
    KOREL is a {\sc FORTRAN} based code that 
    fits time series of observed spectra of a multiple stellar system 
    in the Fourier wavelength domain, 
    to decompose
    them into mean spectra of each component, and simultaneously finds the
    best estimate of the orbital parameters (Hadrava 1995). It also calculates
    RVs of all components by fitting each exposure as a superposition of the
    disentangled spectra.
    We use the recent version
    KOREL08 to find RVs with a sub-pixel precision (cf. Hadrava 2009). 
 
    As the spectral lines are broad and hence more likely
    blended with nearby lines, we carefully chose several wavelength regions
    where the orbital motion could be clearly observed.
    We have chosen regions
    containing different numbers of 
    spectral lines: 
    4537--4595\AA, 4931--4995\AA, 4975--5038\AA,
    5300--5368\AA, 5506--5577\AA, and 6340--6357\AA.
    We only used the data from FEROS, FIES, and FOCES because these provided
    the highest SNR as well as allowing sufficient spectral resolution for 
    the combined data set.

    Each spectral region was disentangled independently converging the
    following parameters: 
    periastron epoch, eccentricity $e$, longitude of periastron $\Omega$, 
    semi-amplitude of the radial velocity curve of the primary component $K_A$,
    and mass ratio 
    $q\equiv K_A/K_B$ (=$M_{B}/M_{A}$).
    The orbital period $\Pi$ 
    was held fixed to the constant 5.71098 days (MR05), 
    because the spectral data can not improve this
    previous determination.
    Also, note that the systemic velocity $\gamma$ is not obtained from 
    disentangling, because KOREL does not use any spectrum template. 
    To determine the optimal orbital parameters, we inspected
    both the {\it residual value} of the KOREL fit as well as the
    extracted RV curves.   However, broad spectral profiles 
    enlarge the range of acceptable orbital values considered as 
    ``good fits''.
    For each of the spectral regions we obtain a set of RV measurements.

    Fig.~\ref{somedisp}  
    shows the observed spectra (circles)
    for the spectral region 4546--4585 \AA\ 
    at orbital phases 0.98 (eclipse) and 0.1 (separated).
    The solid curve is the sum of the individual disentangled primary and 
    secondary spectra, but 
    x-shifted by their 
    RVs at the appropriate orbital phase and, for maintaining clarity, we also 
    vertically shift the phase = 0.1 spectrum.



    \begin{table*}
      \begin{center}\caption{ Spectroscopic orbital solutions based on 
          fitting the 
          radial velocities obtained using different
          data fitting methods, explained in Sect.~\ref{sec:rvs}. 
          \label{tab:solution}}
        \begin{tabular}{lrrrrrrrrr}
          \hline\hline
          & LSD+GAU & LSD+MM & IRAF$_{\rm full}$ & IRAF 
          & KOREL & I09 \\
          &\\

          $e$      &0.27 (0.02)  &0.28 (0.04)  &0.27 (0.08)  
          &0.27 (0.05)  &0.29 (0.01)  & 0.25 (0.04) \\
          $\Omega$ ($^{\circ}$) &83.6 (2.0)   &78.6 (4.9)   &80.5 (9.9) 
          &81.1 (6.0)   &76.8 (5.0) & 46.7 (7.6)  \\
          $T_0$ (days)    &--0.07 (0.05) &--0.06 (0.17) &--0.07 (0.41) 
          &--0.12 (0.22) &-0.1 (0.2) & --0.44 (0.27) \\
          $\gamma$ (\kms)&--28.1 (0.4) &--30.0 (1.5) &--29.2 (2.8) 
          &--29.2 (1.7) & --  & --27.8 (2.3)\\
          $K_A$ (\kms)   & 88.2 (0.9) &84.3 (2.7)    &84.3  (5.1) 
          &91.9 (3.2)  &87.3 (3.5)  & 85.5 (3.9)  \\
          $K_B$ (\kms)   & 94.6 (1.1) &90.7 (2.8) &91.2 (5.6) 
          &100.5 (3.8)&93.0 (4.0) & 95.9 (4.5) \\
          M$_A\sin^3 i$ (\msol) & 1.67 (0.12) &1.45 (0.11) &1.48 (0.23) 
          &1.96 (0.19) &1.58 (0.15) &1.69 (0.18)\\
          M$_B \sin^3 i$ (\msol)&1.55  (0.11) &1.35 (0.11) &1.37 (0.21) 
          & 1.79 (0.16) &1.48 (0.13) &1.51 (0.16) \\
          $\chi^2_R$ &  2.7   &  & 0.04 & 0.07& 0.86 &0.7\\
          \hline\hline
        \end{tabular}
      \end{center}

    \end{table*}



    \section{Orbital parameters based on RVs\label{sec:orbital}}

    From each of the methods explained in Sect.~\ref{sec:rvs} we obtain a {\it set}
    of RV measurements.
    In order to obtain the orbital  parameters of the binary system we
    fit
    each set separately
    using the standard radial velocity
    equations.  The results are summarised in Table~\ref{tab:solution},
    where the orbital period is fixed at 5.70198 days (MR05), 
    $T_0$ is the 
          offset in days from the defined primary minimum, and 
$\chi^2_R$ is the reduced $\chi^2$ value.  For  the LSD+MM method 
we have SNR values instead of observational error measurements, so the $\chi^2_R$ value is not comparable with the other methods.
    The uncertainties quoted are the standard
    formal uncertainties calculated from the fitted parameters and using the 
    observational errors given for each RV data point.
    The results for KOREL are obtained by weight 
    combining all of the RV measurements
    from the individual spectral profiles, and subsequently fitting using the
    standard equations, while the uncertainties reflect the variation in the 
    fitted orbital parameters from the different spectral regions.
    These are the RV data that were later used for the simultaneous
    photometric and spectroscopic analysis (Sect.~\ref{sec:photo}).
    We also analysed the published RVs from I09, who had used standard
    IRAF procedures to determine these.  The results are given under heading 
    I09.

    We define the primary component as the more massive star.
The primary is the star that shows the deeper LSD or CCF profiles as well
    as the line-profile variations due to pulsations 
    (see Sect.~\ref{sec:discussion}).

\begin{figure}
      \includegraphics[width=0.5\textwidth]{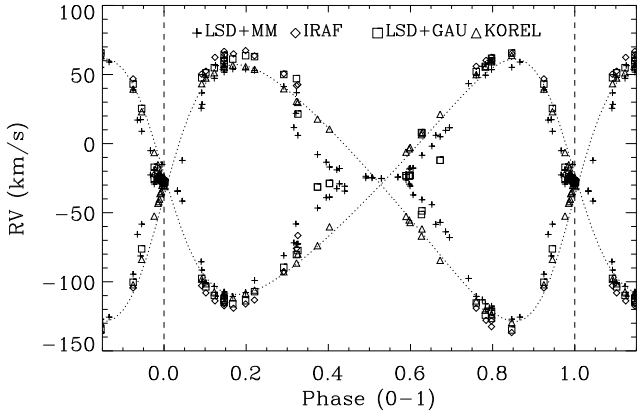}
      \caption{Phased radial velocity solution (dotted curve) obtained 
        using the 
        parameters from Table~\ref{tab:solution} under the column
        heading KOREL.
        The RVs obtained by the individual methods are also shown.
        The dashed lines delimit orbital phase at 0 and 1.
        \label{fig:rvcurve}}
    \end{figure}

   Fig.~\ref{fig:rvcurve} shows the phased radial velocity data 
    from each of the methods: LSD+MM (crosses), IRAF (diamonds), 
    LSD+GAU (squares), 
    and KOREL (triangles --- we have included a weighted average 
    $\gamma$ shift of 
    -28.02 \kms\ for clarity).  
    The dotted curve shows the radial velocity solution
    given by the parameters using KOREL and shifting it vertically by $\gamma$.

    All of the methods presented return values of $e$, $\Omega$, $T_0$ and 
    $\gamma$ consistent with each other (Table~\ref{tab:solution}).  
    There is, however, a variation among the fitted values 
    of $K_A$ and $K_B$.  This is due to the sensitivity of the 
    various methods used
    to extract the radial velocities.
    Two possible reasons that affect the determination of the RVs
    are (1) the effect of the pulsations on the spectral profiles, and 
    (2) line-broadening due to rotation making it difficult to accurately 
    determine RVs (especially at conjunction).

    If we inspect Fig.~\ref{fig:lsdprofile} we see that the double-Gaussian 
    function cannot correctly reproduce the shape of the LSD profile. In fact, 
    due to the {\it wiggles} or {\it variations} in the line, the actual center of the Gaussian 
    could be shifted slightly to the right or the left.
    We have highlighted the center positions derived using the 
LSD+GAU and LSD+MM methods with
    the continuous and dashed vertical lines.  
    In this figure it can also be seen that the latter method is sensitive to the minima
    of the deepest variations in both the primary and secondary profiles.
    In this case, it results in smaller absolute values of the RVs than the 
    former method.  By inspecting these profiles by eye, it is very difficult
    to distinguish which is the true orbital motion RV. 
    
    For phases between $\sim$0.35 and 0.7 we begin to see 
    blending of the component profiles using both CCFs and LSD methods. 
    Hence
    the resulting RVs between these phases will not be as accurate.
    For example, the deviations of the RVs using the LSD+MM method 
    from the fitted RV curve in Fig.~\ref{fig:rvcurve} 
    are explained most likely by 
    the sensitivity of this method to the deformations in the line profiles
    induced by pulsations. 
    
    In Fig.~\ref{fig:ccfpedro} we show the CCF calculated using IRAF procedures
    at two different orbital phases.
    For phase = 0.76 we have also plotted the Gaussian fits.
    Again, it is difficult to extract an accurate RV value 
    because the lines are so broad and due to
    the variations (wiggles) in the CCFs.
    These values were obtained using the upper 50\% of the CCF (although
    we draw the full analytical profile).
    
    By inspecting the profiles of all of the available spectra (and CCFs or
    LSDs) it is not possible to distinguish between shifts 
    of a few \kms, and it is these shifts that cause the differences
    in the fitted values of $K_A$ and $K_B$ of $\sim$6 \kms\ shown in Table~\ref{tab:solution}.

    The last method used to extract the RVs was KOREL.
    We choose this as the best solution because it is robust 
    against broadened/blended lines, it simultaneously fits
    the orbital solution (comparing all of the spectra at the same time), and 
    it is independent of spectral templates.
    These are the RVs that are subsequently used for the simultaneous 
    photometric and spectroscopic light curve fitting.


    \section{Combined analysis of photometric and spectroscopic data\label{sec:photo}}

    In order to obtain a consistent photometric and RV solution for this
    system, we combine the RV data with an extensive photometric data set (MR05).
    We include photometry
    from some more recent (2005 and 2007) observational campaigns with 
    the twin Danish 6-channel $uvby\beta$ photometer at the 1.5m and the 0.9m
    telescopes at 
    San Pedro
    Mart\'{\i}r, and Sierra Nevada Observatories (data reduced in the same 
    manner as MR05).
    We also include photometric data from the P7 photometer mounted
    on the Flemish Mercator telescope at the Roque de los Muchachos Observatory
    (see Rufener 1964, 1985 for reduction procedure).
    In total there are more than 4000 photometric measurements obtained in each of the four
    Str\"omgren filters, spanning more than 3700 days, and 70 out-of-eclipse 
    data points  in the Geneva UBB$_1$B$_2$VV$_1$ filter system.
    We used PHOEBE
    (PHysics Of Eclipsing BinariEs) (version 0.31a; Pr\v{s}a \& Zwitter 2005) 
    and FOTEL (Hadrava 1990) to model these data, however the latter was used
    mainly to confirm the results.
    PHOEBE is a user-friendly 
    software based on the Wilson-Devinney (WD) code (2007 v. of 
    Wilson \& Devinney 1971).
  
    \subsection{Photometric Indices}
    Using out-of-eclipse measurements, we calculated the 
    Str\"omgren indices. We obtain $(b-y) = 0.262$~mag, 
    $m1 =0.134$ mag, and $c1=1.058$ mag.
    The de-reddenned indices were obtained using the method described in
    Philip et al. (1976), 
    these are $E(b-y)$~=~0.158~mag, $(b-y)_0$~=~0.104~mag, 
    $m_0$~=~0.184~mag, and $c_0$~=~1.026 mag.
    If we compare these values to Crawford (1979) we obtain spectral type
    A6V or A7V.  
    These values correspond to a single star with 
    a mass range of between 1.7 and 1.9 \msol\, and a \teff\ of between 7400--8100 K.
    The value of $(b-y)_0$ during the primary eclipse is 0.105 mag, practically 
    the same value as the out-of-eclipse value.
    This implies that the temperatures of both stars are very similar, consistent
    with the estimated effective temperature ratio ($T_B/T_A$) of 1.05 from
    MR05.

    \subsection{PHOEBE}
    We started our fitting by varying
    $T_{\rm eff B}$ between = 8000 and 8250 K.
     The value of $T_{\rm eff A} = 7500$ (or 7750) K is fixed, according to the 
    effective temperature ratio of 1.05 (MR05) (and in
    agreement with the results 
from Sect.~\ref{sec:teff}). 
    We also adopted as starting spectroscopic values those 
    given in Table~\ref{tab:solution} and
    the results from MR05 ($e$ = 0.24, $i$ = 73$^{\circ}$, and $q$ = 0.90).
     We searched for the best solution by using iteratively as free parameters: 
    the luminosity of the primary 
    component 
    $L_A$,  
    separation $a$, $q$, 
    $e$, $\Omega$, 
    $i$, and $\gamma$. 
    The radii, masses, and secondary luminosity are subsequently derived from
    these values.
  
    We modelled the stars as a
    detached system 
    assuming a radiative albedo for both stars 
    equal to 1 and 
    gravity darkening values of g$_A$ = g$_B$ = 1.
    We also attempted to model the data using the semi-detached configuration,
    however, most of the parameters did not converge.
    The 
    model parameters from the best fit are given in Table~\ref{tab:results}.
    The errors are derived using 
        PHOEBE.  The velocities \vsini\ are determined from spectroscopy
          (see Sect.~\ref{sec:abund}) allowing the derivation of the 
          rotational periods $P_{\rm rot}$.


    \begin{table}
      \caption{ Component and orbital results from 
        simultaneous photometric and velocity light 
        curve fitting with PHOEBE. 
          \label{tab:results}}
      \begin{center}
        \begin{tabular}{lllllccc}
          \hline \hline
          & Primary & Secondary \\
          \hline
         M (M$_{\odot}$) & 1.78 (0.24) & 1.70 (0.22) \\
          L (L$_{\odot}$) & 52.42 (2.85) & 22.23 (1.33) \\
          \teff (K) & 7,750  & 8,250  \\
          R (R$_{\odot}$) & 4.03 (0.11) & 2.37 (0.07) \\
           \logg & 3.48 (0.08)& 3.92 (0.08)\\
           \vsini\ (\kms) & 78 (3) & 74  (4)\\
           $P_{\rm rot}$ (days) & 2.50 (0.10) & 1.55 (0.09)\\
          \\
           $e$ & 0.283 (0.001)\\
           $\Omega (^{\circ})$ & 90.6 (0.06)\\
           $a$ (\rsol) & 20.32 (0.06)\\
           $q = K_A/K_B$ & 0.96 (0.001)\\
           $i$ ($^{o}$) & 73.2 (0.6)\\

           \hline\hline
        \end{tabular}
      \end{center}
  \end{table}



    \section{Disentangling of spectra \label{sec:korel}}
    The disentangled spectra were obtained by imposing the orbital solution 
    from KOREL
    shown in Table~\ref{tab:solution}.
    The used version of {\sc KOREL} is limited in resolution to 4000 points.
    However, in most cases it was quicker and more stable to 
    work on 2000 points at a
    time, over wavelength regions of about 100 \AA.
    For this we had to rebin the spectra at the appropriate resolution and 
    carefully choose the endpoints of the regions where there was continuum,
    while also ensuring that overlapping spectral regions are present to 
    facilitate the merging of all of the wavelength regions.
    
    We attempted to disentangle the spectra in the 
    wavelength range from 4600 to 5800 \AA\ because this region is most  
    sensitive to abundances. 
    However, we only obtained reliable results for $\sim$100\AA-length 
    spectral regions
    between 4976--5627 \AA.  
    Unfortunately, this implies that a spectroscopic determination of 
    \teff\ from the Balmer lines is not 
    possible {\it from the disentangled spectra}:
    a better orbital phase coverage and higher SNR spectra are needed in
    order 
    to disentangle 
    these regions correctly.

    Spectra from binary systems 
    normalised to 1 contain no information about the 
    relative light contributions of each of the individual spectra.
    We used the literature values from  
    I09 ($L_{A}$ = 0.61 and $L_B$ = 0.39 where $L_A + L_B = 1$), 
    and the fitted luminosities given by our composite 
    photometric and spectroscopic
    analysis ($L_{A}$ = 0.69 and $L_B$ = 0.31, Sect.~\ref{sec:photo}) 
    to obtain the fractional contributions 
    to the total light.
    We shifted the output spectra to continuum 0, scaled the individual 
    spectra by their corresponding fractional luminosity, and 
    shifted the spectra again to 1 to obtain the disentangled 
    individual spectra.

   \section{Spectroscopic determination of \teff
     \label{sec:teff}}
   Once the spectra were disentangled, we proceeded to obtain the LSD profiles of the 
   individual disentangled spectra (Sect.~\ref{subsec:lsd}), in order to estimate 
   $T_{\rm eff}$ and $\log g$.
   We calculated the LSD profiles using a range of templates spanning 
   $T_{\rm eff}$ (7250--8500 K) and $\log g$ (3.5--4.5), and identified 
   the best values as those that produced the best LSD profiles (smoothest 
   and most symmetric).
   Using the luminosity ratios obtained, 
   $T_{\rm eff}$ of the primary was best fit with a template
   spectrum of  7500 K 
    with $\log g = 4.0$.
   The secondary was best fit with a hotter temperature of 8000 -- 8250 K 
   and $\log g$ = 3.5, although $\log g = 4.0$ was also a good fit.
   In general, the results are much less sensitive to the value of $\log g$
than to changes in \teff.
   
   As the disentangled renormalised spectra depend on the assumed luminosity
   ratio of both components, we performed a few tests 
   to investigate how these results depend on this
   imposed value.  
   We scaled each of the disentangled spectra arbitrarily, and 
   proceeded to 
   identify the best \teff\
   for each of the ``synthetic'' spectra in the same manner as above
   (as double-blind tests). 
    The results were again quite insensitive to the value of $\log g$, while
    the derived values of \teff\ were consistent with results of 7500--7750 K for the primary, 
   and 8000--8250 K for the secondary.  
   This test showed that the identified \teff\ were not obtained from 
   the imposed luminosity ratio of the stars,  
   but rather the shape of the spectra. 
    MR05 obtained an effective temperature ratio 
    ($T_B/T_A$) of 
    $\sim$~1.05 (secondary/primary), in agreement with our result, 
    while I09 obtained a hotter temperature for the primary component 
      than for the secondary ($T_B/T_A \sim 0.96$).

\begin{figure}
      \includegraphics[width=0.5\textwidth]{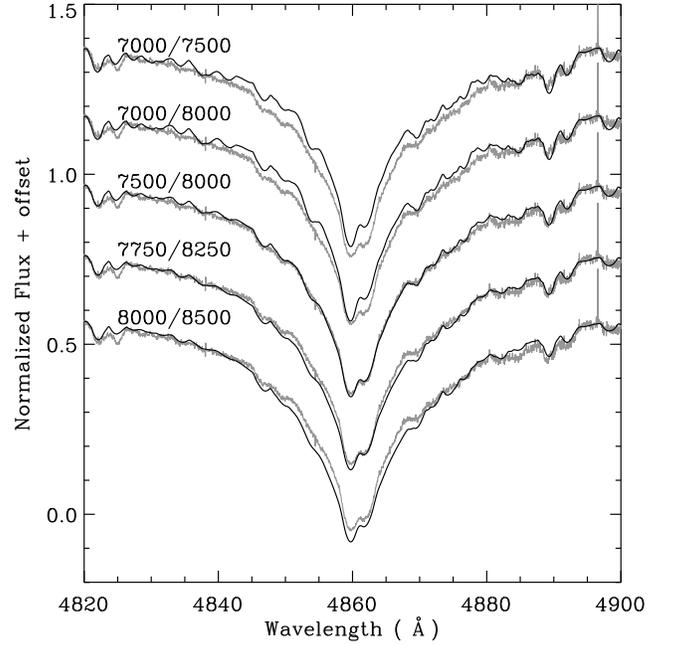}
      \caption{ Composite rotationally broadened RV- and $\gamma$-shifted
        spectral templates (black) with observations (grey) for 
        various combinations of \teff\ (K) for the primary and secondary components.
        \label{fig:teffspec}}
    \end{figure}

    We also directly compared composite rotationally-broadened RV- and 
    $\gamma$-shifted Kurucz synthetic spectra to the best SNR observations 
    near maximum elongation 
    phase.
    We used templates spanning temperatures of 7000 - 8500 K and \logg\ of
    3.5 - 4.5 dex. 
    Figure \ref{fig:teffspec} shows the observations (grey) compared to various 
    templates (black) for the temperature-sensitive H-$\beta$ line.
    As shown in this figure, the best fitted temperatures are
    7500-7750 and 8000-8250 K for primary and secondary star, respectively.
    (The \logg\ values used are 3.5).

\section{Abundance Analysis\label{sec:abund}}


    \begin{figure}
      \includegraphics[width=0.5\textwidth]{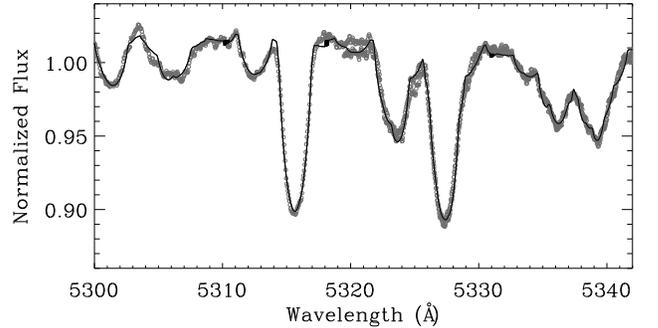}
      \caption{ The disentangled observed primary spectrum (open circles), and 
        the theoretical fitted spectrum (solid) obtained using the 
        abundances from Table~\ref{tab:abund}. 
        \label{fig:fitspec2}}
    \end{figure}
    
    \begin{table}
      \caption{ The abundances of chemical elements for the 
        primary and secondary stars.\label{tab:abund}}
      \begin{center}
        \begin{tabular}{lcllcc}
          \hline\hline
          Atomic& No. lines used & Primary & Secondary & Sun\\
          No.& Prim./Sec.\\ 
          \hline
          6 & 4/2  &8.45 (0.15)&  8.78 (0.28)&8.39\\
          8 & 1/1  &8.89 (0.13)&  8.82 (0.13)&8.66\\
          11 & 1/--  &6.21 (0.08)&            &6.17\\
          12&2/2 &7.35 (0.16)&  8.37 (0.30)&7.53\\
          13&1/-- &6.17 (0.38) &  &6.37 \\
          14&2/2 &7.81 (0.21)&  7.91 (0.12)&7.51\\
          16&1/-- &6.67 (0.21)&             &7.14\\
          20&4/4 &6.28 (0.18)&  6.92 (0.23)&6.31\\
          21&2/2 &3.12 (0.10)&  3.89 (0.26)&3.17\\
          22&9/9 &4.90 (0.18)&  5.38 (0.21)&4.90\\
          24&4/5 &5.81 (0.14)&  6.31 (0.13)&5.64\\
          26&19/20 &7.17 (0.21)&  7.84 (0.19)&7.45\\
          27&1/-- &5.57 (0.02)&  &4.92\\
          28&6/9 &6.13 (0.18)&  6.89 (0.22)&6.23\\
          39&2/2 &2.72 (0.20)&  2.78 (0.17)&2.21\\
          40&--/1 &           &  3.20 (0.08)&2.58\\
 
            
          \hline\hline
          
        \end{tabular}
      \end{center}
    \end{table}


    We determined the abundances of both components using 
    the disentangled renormalised spectra.
    We compared the results with both values of luminosity fraction, 
    and the final abundances varied 
    slightly, but within the quoted errors.
    Our analysis follows the methodology presented in 
    Niemczura \& Po{\l}ubek (2006)
    and 
    relies on an efficient spectral synthesis based on
    a least-squares optimization algorithm (Bevington 1969, Takeda 1995).
    This method allows the simultaneous
    determination of various parameters associated with stellar spectra and
    consists of minimizing the deviation between the theoretical flux
    distribution and the observed one.

    The atmospheric models used for the synthethic spectra determination
    were computed with the line-blanketed LTE ATLAS9 code
    (Kurucz 1993), which handles line opacity with the 
    opacity distribution function.
    The synthetic spectra were computed with the SYNTHE code (Kurucz 1993).
    Both  codes were ported to GNU/Linux by Sbordone (2005)\footnote{
      These are available online: {\tt wwwuser.oat.ts.astro.it/atmos/}}.
    The identification of stellar lines and the abundance analysis 
    were performed on a line list we constructed based on 
    VALD\footnote{{\tt http://ams.astro.univie.ac.at/\~{}vald/}} 
    data (Kupka et al. 2000 and references therein).
    
    The shape of the synthetic spectrum depends on the
    stellar parameters such as \teff, \logg, \vsini, 
    microturbulence \turb, RV 
    and relative abundances of the elements.
    The effective temperature and surface gravity were not determined 
    during the iteration
    process but were considered as fixed input ones
    (T$_{\rm eff}$ = 7500/8000 and \logg = 4.0/4.0 for primary/secondary 
    components),
    while \turb\ = 2\,\kms\ 
    was adopted. 
    All of the other above-mentioned parameters can be determined 
    simultaneously because they produce detectable and different
    spectral signatures.
    The theoretical spectrum was fitted to the normalised observed one in 
    small regions, until the results converged.
    The obtained chemical element abundances, relevant errors with the 
    adopted solar abundances
    (Grevesse et al. 2007) are given 
    in Table~\ref{tab:abund}.
    The errors are calculated as
    $\sigma_{\rm Tot} = [\sigma_{\rm ab}^2 + \sigma_{T_{\rm eff}}^2 + 
      \sigma_{\log g}^2]^{1/2}$ where $\sigma_{\rm ab}$ is the 
    standard deviation of the individual element abundance, 
    $\sigma_{T_{\rm eff}}$ and $\sigma_{\log g}$ denote the 
    variation of abundance for $\Delta T_{\rm eff} \pm 250$ K and 
    $\Delta \log g \pm 0.2$.
        For those elements where only one line was used to estimate
        the abundance, the 
        error is $(\sigma_{T_{\rm eff}}^2 + \sigma_{\log g}^2)^{1/2}$.

    Apart from the errors in $T_{\rm eff}$ and \logg, several systematic 
    errors 
    are likely affecting the derived abundances.
    For example, fixing \turb, the restriction of the disentangling 
      technique
      to the spectral range of 4976--5672 \AA\, 
      and the imposed luminosity 
      ratio all contribute to these sytematic errors.
    In addition to this, 
    the adopted atmospheric models and the 
    atomic data (in particular the oscillator strengths) 
    also contribute.
    We are also not taking NLTE effects into account, which are
    crucial for a number of spectral lines of several elements 
    (see e.g. Asplund 2005; Fabbian et al. 2006, 2009).
    However, we have estimated that for the neutral C lines used here, 
    the NLTE corrections are negligible (private comm. Fabbian). 
    Moreover, even for other chemical elements we can safely assume that 
    the differential NLTE effects between the primary and secondary 
    stars are 
    small and hence can not explain the differences for some of the 
    elements between the two stars.

    Figure \ref{fig:fitspec2} shows a region of the primary disentangled
    spectrum (circles) and 
    the theoretical fitting to it (solid curve)
    which is obtained using the abundances from Table~\ref{tab:abund}. 
    The chemical composition of the primary star is slightly 
    sub-solar, judging
    from the Fe content, while 
    the secondary star shows enhanced abundances 
    (see Fig.~\ref{fig:abund}).
    The derived rotational velocities 
    are \vsini\ = 78 $\pm$ 3 \kms
    and 74 $\pm$ 4 \kms for primary and secondary component respectively, 
    based on a comparison between 
    the spectra with broadened
    theoretical profiles.
    Coupling these values with the results for \rsol\ and $i$ 
      from Table~\ref{tab:results}, we 
      derive the following rotational periods: 
      $P_{\rm rot,A} = 2.50 \pm 0.10$ days for the primary star 
      and $P_{\rm rot,B} = 1.55 \pm 0.09$ days for the secondary.

    We have also performed a comparison abundance analysis based on a detailed
    line abundance approach using the MOOG\footnote{MOOG was developed 
      by Chris Sneden, see: http://verdi.as.utexas.edu/moog.html} code.
    The full procedure is described in Arellano Ferro et al. (2001) and 
    Giridhar \& Arellano Ferro (2005).
    The number of elements analysed is smaller than those shown in 
    Table~\ref{tab:abund}, but the overall Fe abundance and the indication that
    the secondary component is iron rich is consistent with our previous result.
    This analysis yielded the best results when \teff\ of 7750 and 8000 K,
     and \logg\ of 3.0 and 3.5
    were adopted for primary and secondary, respectively.

    \section{Discussion\label{sec:discussion}}
  
    \subsection{Orbital and System parameters}

    From the spectroscopic and the combined photometric and spectroscopic 
    analysis we obtain system parameters that are in agreement:
    $e = 0.29$, $\Omega \sim 80^{\circ}$ and 
    $T_0 \simeq -0.1$ (this result is compatible with 
    $T_0 \simeq 0.0$ taking the error into account, and this value 
    indicates that the resulting offset in days from eclipse/conjunction
    coincides with the input value from MR05).  
    When we compare our results with the only other published system
    parameters by I09, we find that I09 obtain from their combined 
    photometric and spectroscopic analysis a lower
    value of the eccentricity (e=0.19).
    When we fit their data using our RV
    solution method, we obtain $e = 0.25$, $\Omega \simeq 47^{\circ}$ and 
    $T_0 = -0.4$.  However, when we use $\Omega=80^{\circ}$ and $T_0=-0.1$
    as fixed values we obtain  
    a $\chi^2$ value that increases 
    by less than 1 (i.e. less than 1$\sigma$), 
    indicating that this solution is also plausible with their 
    data.
    The average $\gamma = -28.0$ \kms\ is consistent with the cluster radial 
    velocity of --25.0 \kms (Valitova et al. 1990): providing extra
    evidence of the system's membership to the cluster.

    The derived value of $i = 73^{\circ}$ indicates that 
    if we aim to determine precise masses, we must accurately 
    determine the values of $K_A$ and $K_B$.
    Given that we obtain a range of about 6 \kms\ in both $K_A$ and $K_B$
    this implies that the minimum mass error is already on the order of 
    a few tenths (if we just use the analytical calculation and the inferred 
    inclination).
    The resulting values for $M\sin^3 i$ are listed in Table~\ref{tab:solution},
    where it can be seen that the variation is of the order of 0.3 \msol\ 
    (all however within 3$\sigma$ of each other).
    In Sect.~\ref{sec:orbital} we already  discussed the sensitivity of
    each of the methods in obtaining the RVs, and we would like to 
    point out the difficulty of attempting to model 
    the type of stellar system considered here.
    What hampers the precise determination of 
    the RVs of each component is the 
    significant broadening
    of the spectral line profiles due to the rotation of 
    both stellar components, and the superimposed pulsations.

    The largest differences in the results come from using the same method
    (IRAF) but choosing a different cut-off region in the CCF profiles to model.
    While the two-Gaussian function did not
    show reasonable results for IRAF$_{\rm FULL}$, 
    an adequate 3-Gaussian fitting to the CCF seemed
    to fit the profile
    better indeed.  
    This third low amplitude component could be a sign of a 
    third body in the orbit. 

    When combining the photometric and spectroscopic data (Sect.\ref{sec:photo})
    it was difficult to arrive at a solution that (iteratively) converged 
    using the parameters  
      listed in Table~\ref{tab:results} as free parameters. 
    This is most likely a result of the rapid rotation causing surface 
    temperature variations at different latitudes 
    (sometimes a difference of up to
    1,500 K for very rapid rotation).
    The photometric light curve is distorted due to these variations, 
    and these effects are not taken into account in the 
    modeling (the assumption with PHOEBE is that
    the star rotates as a rigid body, without differential 
    rotation\footnote{See the PHOEBE manual, available at {\tt http://phoebe.fiz.uni-lj.si/} }). 
    Additionally, one of the components has visible photometric variability
    on time scales of $\sim$1 hour 
    (MR05, Amado et al. 2006, Costa et al. 2007), 
    and although the light curves were binned to 
    reduce the effects of pulsation, there are nonetheless effects that can 
    not be eliminated, such as those during eclipse.
    
    We obtain component masses 
    of 1.78 $\pm$ 0.24 and $1.70$ $\pm$ 0.22 \msol\
    for primary and secondary stars respectively, 
    values that are lower by $\sim0.2$ \msol\ than those from 
    I09 (2.06 and 1.87
    \msol).
    Our results are consistent with the values obtained by 
    Mermillod \& Mayor (1990) who, through a study of red giants, determined
    the main sequence (MS) turn-off mass of IC~4756 to be 
    between 1.8 and 1.9 \msol.
    This would also be consistent with the hypothesis of the more massive
    component beginning to turn off the MS, while also implying that
      the secondary star is slightly 
    hotter than the primary. 

    We obtain primary and secondary 
    radius of 4.03 $\pm$ 0.11 
    and 2.37 $\pm$ 0.07 \rsol, respectively,
     luminosities of 52.42 $\pm$ 2.85 and 22.23 $\pm$ 1.33 \lsol, 
    and \logg\ values of
    3.48$\pm$0.08 and 3.92$\pm$0.08 (see Table~\ref{tab:results}).
    These values are lower than those quoted by I09, due primarily
    to the lower fitted masses we obtain.

    We finally fixed the values of \teff\
    at 7500--7750 and 8000--8250 K respectively.
    This choice is justified by the following reasons:
    (a) at an earlier stage during this work, modeling the photometric 
    light curve produced similar answers (7,560 and 8,030 K) while fixing 
    various other parameters, 
    (b) modeling the light curve while fixing the \teff\ at other values
    results in worse fits or non-convergence, 
    (c) the photometric Str\"omgren indices are consistent with 
    these ranges of values 
    (Sect.~\ref{sec:photo}), and
    (d) the spectroscopic determination of \teff\ from the LSD profiles 
    and synthetic spectra also arrives at these values (Sect.\ref{sec:teff}).

    The primary (hotter and more massive) star is the pulsating
      component that clearly shows line-profile variations.
      Our data do not allow us to determine if the secondary 
      component is also pulsating
      as the SNR is too low to detect low-amplitude variability, if any.
      The results in Sect.~\ref{sec:photo} and Table~\ref{tab:results} do not
      discard the possibility of two pulsating components ---
      both stars lie within the $\delta$ Scuti instability strip close
      to the blue edge (see Fig.~13 in Pamyatnykh 2000).

    We have explored
    various forms of determining the RVs from the 70 spectra collected
    between 2005 and 2007 as well as using photometric observations spanning
    over 3000 days.  
    Taking into account the uncertainties, our results are consistent
      with the I09 ones, however, apart from the RV dataset derived with 
      the IRAF method, we obtained lower 
      masses and this would imply a downward
    revision of these values to 
    1.8 and 1.7 \msol.

    \subsection{Abundances}
 
    The resulting abundances given in Table~\ref{tab:abund} 
    indicate for the primary component a sub-solar metal value of 
    [Fe/H]~$=-0.28$ 
    consistent with the values of --0.15 and --0.22 determined for the 
    cluster IC~4756 derived from samples of F-G single 
    star cluster members by Jacobson et al. 
    (2007) and Thogersen et al. (1993), respectively.
    The results of I09 also showed that this system component's global parameters
    fitted better
    to a sub-solar metallicity evolutionary track ($Z\sim0.008$, where  $Z$ is the 
    initial metal mass fraction).
    For HD~172189  we obtain the following abundances (primary/secondary):
    [Na/Fe]~=~+0.03/--, [Al/Fe]~=~+0.00/--, 
    [Si/Fe]~=~+0.08/+0.00, 
    [Ca/Fe]~=~+0.03/+0.04, and [Ni/Fe]~=~+0.02/+0.04.
    These values are consistent with those of IC~4756 according to
    Jacobson et al. (2007).  However, their values are higher than ours by up to
    $\sim$0.3 dex (they quote +0.6, +0.3, +0.34, +0.07, +0.08, respectively).
    This could be in part due to the adopted $\log-gf$ values 
    (Jacobson et al. 2007).
   
    \begin{figure}
      \center{\includegraphics[width = 0.5\textwidth]{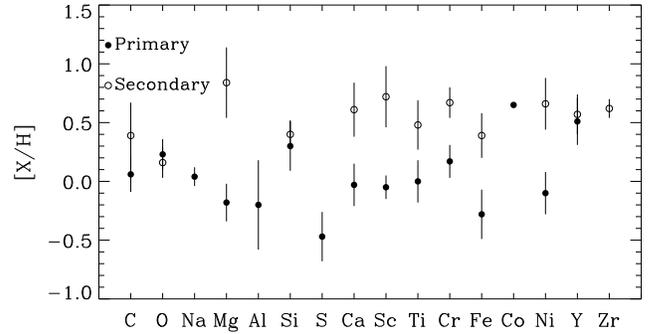}}
      \caption{ Abundances of the primary (filled circles) and 
        secondary (open circles) stars 
        compared to the adopted solar values.
        The bars represent errors of $\pm 1\sigma$.\label{fig:abund}}
    \end{figure}

    The abundances relative to hydrogen for the primary star are:
    $[{\rm Fe/H}]=-0.28$, [Si/H]~$=+0.3$, [Ca/H]~$=-0.03$, [Ni/H]~$=-0.10$, 
    [Na/H]~$=+0.04$. 
    These values are comparable to Jacobson et al. (2007), who obtained
    --0.15, +0.19, --0.08, --0.07, 
    +0.2\footnote{Here, Jacobson et al. used the $gf$ values from 
      Luck (1994).} for IC~4756.
    The results for the secondary star are
    [Fe/H]~=~+0.4, [Si/H]~=~+0.4, [Ca/H]~=~+0.6, [Ni/H]~=~+0.6,
    on average $\sim$0.5 dex higher than the
    primary star, resulting in 
    enhanced abundances compared to the Sun.    
    Fig.~\ref{fig:abund} shows the abundances 
    relative to the solar ones.
    The enhanced abundances of the secondary star
    could also be an effect of modeling a spectrum 
    with an inadequate normalisation from the disentangling.
    Higher SNR disentangled spectra resulting 
    from the full orbital phase coverage would be able to confirm these
    different abundances.

    \section{Conclusions\label{sec:conclusions}}

    We analysed a set of spectroscopic data 
    obtained between 2005 and 2007 with 
    the aim of determining the absolute component masses, radii, luminosities, 
    and effective temperatures of the system, as well as undertaking the first 
    abundance
    analysis of the HD~172189 disentangled spectra.
    From a combined photometric and radial velocity analysis, we derive
    for the primary and secondary stars, respectively,
    masses of 1.78$\pm$0.24 and 1.70$\pm$0.22 \msol\ 
    radii of 4.03$\pm$0.11 and 2.37$\pm$0.07 \rsol, 
    luminosities of 52.42$\pm$2.85 and 22.23$\pm$1.33 \lsol,
    and \logg\ of 3.48$\pm$0.08 and 3.92$\pm$0.08 dex.
    The analysis of the spectra indicates a \teff\
    of 7600$\pm$150 K and 8100$\pm$150 K.
    From the RV data, we derive the system orbital parameters: 
    $e = 0.29 \pm 0.02$, $\Omega = 77^{\circ}\pm5$ (from KOREL), 
    and $\gamma = -28.02 \pm 0.37$ \kms\ 
    (from other RV methods),
    and from the photometric data $i = 73.2^{\circ}\pm0.6$
    and $a = 20.32\pm0.06$ \rsol.
    The spectroscopic data also 
    confirm the orbital period of the system: 5.70198 days, as derived
    by MR05. 
    We also obtain a spectral type of A6V-A7V based on Str\"omgren photometry.
    
    We applied four methods to calculate
    the RVs and obtained different results for each.
    We subsequently showed the limitations of each technique for analysing
    a system with two rapidly-rotating components, one of these having
    the additional complications of showing significant profile-variations
    due to pulsations. 
    Obtaining higher SNR spectra with continuous
    phase coverage 
    would help to determine the systematic differences between 
    each of the methods.

    We have disentangled the spectra of both components and determined 
    the rotational velocities: \vsini\ = 78$\pm3$ \kms\ and 74$\pm$4 \kms. 
    Coupling these values with the results for \rsol\ and $i$ we obtain
    rotational periods $P_{\rm rot,A} = 2.50 \pm 0.10$ days and
    $P_{\rm rot,B} = 1.55 \pm 0.09$ days.
    We subsequently
    derived a metallicity of [Fe/H] = --0.28 dex 
    and abundances of
    [Si/H] = +0.3, [Ca/H] = --0.03, [Ni/H] = --0.10 and 
    [Na/H] = +0.04 for the primary star.
    These are consistent with the results published for the 
    cluster IC~4756 by Jacobson et al. (2007).
    The sub-solar metallicity is also consistent with the findings from I09. 

    The rapid rotation of both components, the non-synchronous
    rotation (based on the orbital period), the eccentricity, and the
    likely membership of IC 4756 suggests that the  system 
    is still detached. 
    Based on the inferred age of the system of 0.8--0.9 Gyr
    (Alcaino 1965, Mermillod \& Mayor 1990)
    we estimate that the 1.8 \msol\
    primary star is 
    moving towards the end of 
    its MS lifetime, when it begins to cool down and increase in 
    luminosity.
    Mermillod \& Mayor (1990) also determine that the main sequence 
    turn-off mass 
    for this cluster is 1.8--1.9 \msol, which agrees with this hypothesis.

    In order to study the oscillations of the primary star, 
    recently CoRot (Baglin et al 2006a,b, Michel et al. 2008) observed
    HD~172189 as a primary asteroseismic target. 
    Our analysis of this data has given a thorough insight into the nature
    of this object, and hence 
    forms a solid foundation for the subsequent asteroseismic
    analysis.
    We look forward to the unravelling of the mysteries associated 
    with this object, and 
    to learning about 
    the effects of rotation on
    the interior structure of the star from the pulsation frequencies.

    \begin{acknowledgements}
      The FEROS data were obtained with ESO Telescopes at the La Silla 
      Observatory under  the ESO Programme: 075--D.032.
      This work is partially based on observations made with the Nordic 
      Optical Telescope, jointly operated
      on the island of La Palma by Denmark, Finland, Iceland,
      Norway, and Sweden, in the Spanish Observatorio del Roque de los
      Muchachos of the Instituto de Astrofisica de Canarias. 
      We acknowledge Calar Alto director, Joao Alves, for authorising
      schedule changes and thank astronomers F. Hoyo, M. Alises and 
      S. Pedraz and
      the rest of the staff for acquiring the FOCES dataset, 
      at the Centro Astron\'omico Hispano Alem\'an (CAHA) at Calar Alto, 
      operated jointly by the Max-Planck
      Institut f\"ur Astronomie and the Instituto de Astrof\'{\i}sica 
      de Andaluc\'{\i}a (CSIC). 
      Part of this work is based on observations made with the Mercator 
      Telescope, operated on the island of La Palma by the Flemish Community, 
      at the Spanish Observatorio del Roque de los Muchachos of the Instituto 
      de Astrofísica de Canarias,
      and
      observations (SPM and SNO) obtained at the Observatorio de Sierra 
      Nevada (Spain) and at the Observatorio Astron\'omico Nacional San 
      Pedro M\'artir (Mexico).
      OLC is greatful to Petr Hadrava and Ji\v{r}\'{\i} Kub\'{a}t 
      from the Astronomical 
      Institute in Ond\v{r}ejov, Czech Republic, for their hospitality and 
      efforts in making the  
      {\it First Summer School in Disentangling of Spectra} a success.
      SMR acknowledges a "Retorno de Doctores" contract of the Junta 
      de Andaluc\'{i}a.
      PJA acknowledges financial support from a "Ram\'on y Cajal" contract of 
      the Spanish Ministry of Education and Science.
      PH acknowledges the support from a Czech Republic project LC06014.
      CRL acknowledges an \'Angeles Alvari\~no contract under Xunta de Galicia.
      AM ackowledges financial support from a ``Juan de la Cierva'' 
      contract of the Spanish Ministry of Education and Science.
      MR, GC, and EP  acknowledge financial  support from the Italian ASI-ESS
      project,
      contract   ASI/INAF I/015/07/0, WP 03170.
      JCS acknowledges support from the "Consejo Superior de Investigaciones
      Cient\'{\i}ficas" by an I3P contract
      financed by the European Social Fund and from the Spanish
      "Plan Nacional del Espacio" under project ESP2007--65480--C02--01.
      EN acknowledges financial support of the N N203 302635 grant
      from the MNiSW.
      DF acknowledges financial support by the European Commission
      through the Solaire Network (MTRN-CT-2006-035-484).
      We would also like to thank the anonymous 
      referee for their constructive comments and 
      suggestions. 
      This research was motivated and 
      in part supported by the European Helio- and Asteroseismology
      Network (HELAS), a major international collaboration funded by the 
      European Commission's Sixth Framework Programme.

    \end{acknowledgements}


\end{document}